\documentclass[11pt]{article}

\usepackage{graphicx,psfrag,epsf}
\usepackage{enumerate}

\usepackage{url} % not crucial - just used below for the URL 

\usepackage{amsmath}
\usepackage{amsthm}
\usepackage{amssymb}
\usepackage{mathrsfs}

\usepackage{algorithmic}
\usepackage{subcaption}
\usepackage[labelsep=period]{caption}

\usepackage{dsfont} % blackboard numerals
\usepackage[scaled]{helvet}
 
\usepackage[T1]{fontenc}

\usepackage{setspace}
\usepackage{color}
\usepackage{soul}

\newcommand{\E}{\mathbb{E}}

\newtheorem*{defn*}{Definition}

\usepackage[numbers,sort&compress]{natbib}

\usepackage[top=1in, bottom=1in, left=1in, right=1in]{geometry}

\usepackage{tikz}
\usetikzlibrary{shapes,trees,positioning,arrows,matrix,patterns}
\usetikzlibrary{decorations.pathmorphing,decorations.pathreplacing}

\title{Interpretation of the individual effect under treatment spillover}

\author{Forrest W. Crawford$^{1,2,3}$, Olga Morozova$^{1}$, Ashley L. Buchanan$^{4}$, and Donna Spiegelman$^{1,5,6}$ \\[1em]
\normalsize 1. Department of Biostatistics, Yale School of Public Health \\
\normalsize 2. Department of Ecology \& Evolutionary Biology, Yale University \\
\normalsize 3. Yale School of Management \\
\normalsize 4. College of Pharmacy, University of Rhode Island \\
\normalsize 5. Department of Statistics \& Data Science, Yale University \\
\normalsize 6. Center for Methods of Implementation and Prevention Science, Yale School of Public Health}

\begin{document}

%\def\spacingset#1{\renewcommand{\baselinestretch}%
%{#1}\small\normalsize} \spacingset{1}

%%%%%%%%%%%%%%%%%% TITLE %%%%%%%%%%%%%%%%%%%%%%%%%%%%%%%%%%%%

\maketitle

%%%%%%%%%%%%%%%%%% MAIN TEXT %%%%%%%%%%%%%%%%%%%%%%%%%%%%%%%%

Some interventions may include important spillover or dissemination effects between study participants \citep{benjamin2017spillover}. For example, vaccines, cash transfers, and education programs may exert a causal effect on participants beyond those to whom individual treatment is assigned. 
In a recent paper, \citet{buchanan2018assessing} provide a causal definition of the ``individual effect'' of an intervention in networks of people who inject drugs. This work builds on a definition of the ``direct effect'', randomization design, and framework for causal inference under interference introduced by \citet{hudgens2008toward}.  Some researchers have suggested that the ``direct effect'' may not always have a causal interpretation \citep{vanderweele2011effect,savje2017average,eck2018randomization}. In this short note, we discuss the interpretation of the individual effect when a spillover or dissemination effect exists. 

\subsection*{Potential outcomes and causal effects defined by \citet{buchanan2018assessing}}

\citet{buchanan2018assessing} introduce potential outcome notation for the effect of an intervention on an undesirable outcome (e.g. risk behavior, fatal overdose, HIV infection) among people who inject drugs. Let $Y_{ki}$ be an indicator for that outcome, where $k=1,\ldots,K$ is the cluster, and $i=1,\ldots,n_k$ is the individual within that cluster. Let $X_k$ be an indicator that cluster $k$ is treated, meaning that a single cluster member is the ``index'' participant who directly receives the intervention.  
In conformity with the causal inference literature, we use the terms ``treatment'' and ``intervention'' interchangeably.
Let $R_{ki}$ be the indicator that individual $i$ is the index, with exactly one index individual per cluster.  Define the vector of index indicators as $\mathbf{R}_k=(R_{k1},\ldots,R_{kn_k})$.  Define the individual potential outcome $Y_{ki}(\mathbf{r},x)$, where $\mathbf{R}_{k}=\mathbf{r}$ is the vector of index subject indicators and $X_k=x$ is the group-level treatment indicator. Because $\sum_i R_{ki}=1$ for all clusters $k$, the potential outcome notation $Y_{ki}(\mathbf{r},x)$ reduces unambiguously to $Y_{ki}(r,x)$ where $r=R_{ki}$ indicates that individual $i$ is the index; whenever $R_{ki}=1$, it is implicit that $R_{kj}=0$ for $j\neq i$.  \citet{buchanan2018assessing} defined the \emph{individual effect} as 
\[ RD^I = \E[Y_{ki}(1,1) - Y_{ki}(0,1)], \]
the \emph{disseminated effect} as 
\[ RD^D = \E[Y_{ki}(0,1) - Y_{ki}(0,0)], \]
and the \emph{composite effect} as 
\[ RD^{\text{Comp}} = \E[Y_{ki}(1,1) - Y_{ki}(0,0)], \]  
where expectation is with respect to the potential outcomes of individuals $i$ across clusters $k$ in the study. 
The composite effect can be written as the sum of the individual and disseminated effects, $RD^{\text{Comp}} = RD^I + RD^D$.  

\subsection*{Meaning of the individual effect}

\begin{figure}
\centering
	\includegraphics{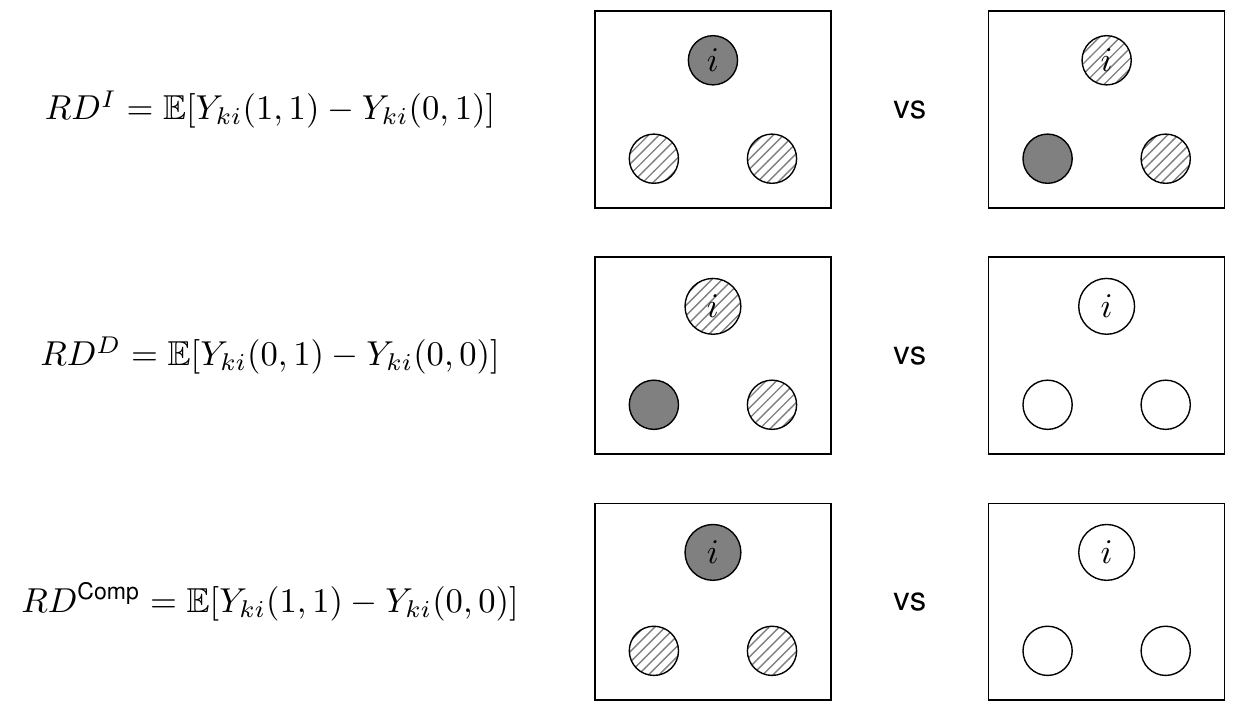}
  \caption{Schematic illustration of the individual, disseminated, and composite effects defined by \citet{buchanan2018assessing} in a cluster of size 3. Cluster $k$ is shown as a rectangle and subjects are shown as circles, with subject $i$ labeled. Gray shading 
  {\protect\tikz[baseline=-0.5ex]{\protect\node[minimum size=0.5cm,inner sep=0.05cm,fill=gray,circle,draw]{};}} \ 
  indicates a treated index subject, and a hatched pattern 
  {\protect\tikz[baseline=-0.5ex]{\protect\node[minimum size=0.5cm,inner sep=0.05cm,pattern=north east lines,pattern color=gray,circle,draw]{};}} \ 
  indicates a subject exposed to treatment via dissemination from a treated cluster member. The individual effect $RD^I$ compares the potential outcome of subject $i$ when treated with no dissemination from group members, versus the potential outcome of subject $i$ when untreated with dissemination from the treated index.}
	\label{fig:exposure}
\end{figure}

The potential outcome notation of \citet{buchanan2018assessing} implicitly encodes two distinct types of exposure to the intervention for subject $i$ in a treated cluster $k$. First, if subject $i$ is the treated index ($R_{ki}=1$, $X_k=1$), then $i$ receives exposure to the intervention via their own treatment, and no disseminated exposure from another cluster member, because no other cluster members can be treated.  Second, if subject $i$ is a non-index in a treated cluster ($R_{ki}=0$, $X_{k}=1$), then $i$ receives no direct exposure to the intervention, but receives disseminated exposure from one treated index subject in their cluster.  Figure \ref{fig:exposure} shows how the individual effect $RD^I$ contrasts potential outcomes by changing both of these types of exposures \textit{simultaneously}. 

\citet[][Table 1, page 2450]{buchanan2018assessing} interpret $RD^I$ as the ``effect on persons directly receiving an intervention beyond being in an intervention network''. 
When the outcome is undesirable and $\E[Y_{ik}(0,0)] \ge \E[Y_{ik}(0,1)] \ge \E[Y_{ik}(1,1)]$,
$RD^I$ may indeed summarize the additional benefit of being an index, ``beyond'' that of experiencing disseminated effect.  
However, this monotonicity relation may not always hold: an intervention with a strong disseminated effect may benefit untreated cluster members to a greater extent than those who personally receive treatment. In this case, the magnitude of the disseminated effect $RD^D$ would be greater than that of the composite effect $RD^{\text{Comp}}$, and the interpretation of $RD^I$ as the additional benefit of being an index, ``beyond'' that of experiencing disseminated effect, may not be meaningful.

\subsection*{A hypothetical intervention trial with a strong dissemination effect}

One of the most effective interventions for reversing a potentially fatal opiate-related overdose is intranasal or injection administration of naloxone during an overdose.  Consider a hypothetical study of clusters of people who inject drugs at risk of fatal opiate overdose, and an intervention that involves dispensing a naloxone kit to one cluster member and training that individual in its administration.  
The mechanism of action of this intervention in groups induces asymmetry in its effects: overdose involves unconsciousness or incapacitation, so naloxone is rarely self-administered; instead, someone who has it and has been trained in its use can avert another cluster member's overdose. 

Suppose that the outcome of interest is fatal overdose, and treatment involves receipt of a naloxone kit from investigators and training in its use.  A subject whose fellow cluster member is treated enjoys a measure of protection against death due to overdose, because the treated subject can use their naloxone kit to reverse their cluster member's overdose, so $RD^D<0$. In contrast, the treated subject may derive less benefit from their own treatment because they cannot use their own naloxone kit to reverse their own overdose, except in rare cases \citep{green2014two}.  It is possible, but perhaps less likely, that their fellow cluster member not trained by investigators in its use might administer the treated subject's naloxone kit to the treated subject, should they experience an overdose, implying $RD^D<RD^{\text{Comp}}\le 0$. 
Clearly treatment is beneficial to any individual whose fellow cluster member receives it, and is either beneficial or ineffective to individuals who only receive it themselves.  However, the individual effect is $RD^I = RD^{\text{Comp}} - RD^D > 0$, so treatment seems to be harmful to the subject who receives it. 
Of course, naloxone is not harmful to anyone in this scenario; rather, the ``individual effect'' contrasts the small (or nonexistent) beneficial effect of individual treatment ($RD^{\text{Comp}})$ against the larger beneficial disseminated effect of treatment ($RD^D$).

\subsection*{Discussion}

The quantity $RD^I$ introduced by \citet{buchanan2018assessing} is a well-defined statistical estimand. However, $RD^I$ may be misleading because it does not measure the effect of the intervention on the individual who would receive it, holding treatment to others constant. Instead, $RD^I$ contrasts the potential outcome of an individual who receives treatment but no disseminated exposure with that of an individual who receives no treatment but disseminated exposure from another subject.  When the disseminated effect is large compared to the composite effect, $RD^I$ can be positive (suggesting harm) even when the intervention is beneficial to treated individuals and their fellow cluster members.  
%, or negative (suggesting benefit) when the intervention is harmful to individuals.   
\citet[][Table 4, page 2455]{buchanan2018assessing} found a similar pattern in their evaluation of the HPTN 037 trial intervention: the individual effect on any risk behavior $RD^I$ is estimated to be null (ineffective), even though the disseminated and composite effects were estimated to be negative (beneficial).  
%\hl{In another example of substantial public health importance, early access to antiretroviral therapy (ART) among HIV positive individuals \mbox{\citep{cohen2011prevention,insight2015initiation,donnell2010heterosexual}}, and pre-exposure prophylaxis among HIV negative individuals \mbox{\citep{baeten2012antiretroviral,thigpen2012antiretroviral,choopanya2013antiretroviral}}, might lead to little or no benefit in terms of longer high quality life among positives or protection from infection among negatives, and may expose those directly treated to significant side effects from these medications.  Spillover to their sexual partners, infants, and overall in their communities could, on the other hand, be substantially beneficial \mbox{\citep{world2016consolidated}}.}
Two additional examples of substantial public health importance could show a similar pattern. Early access to antiretroviral therapy (ART) among HIV positive individuals is known to improve health and helps prevent transmission to HIV-negative partners \mbox{\citep{temprano2015trial,insight2015initiation,cohen2011prevention,donnell2010heterosexual}}. Likewise, pre-exposure prophylaxis (PrEP) among HIV negative individuals helps prevent HIV infection in treated individuals \mbox{\citep{baeten2012antiretroviral,thigpen2012antiretroviral,choopanya2013antiretroviral}}. While these interventions are known to benefit treated individuals, and reduce HIV transmission within groups \mbox{\citep{world2016consolidated}}, a trial that computed an ``individual effect'' by contrasting $RD^{\text{Comp}}$ against $RD^D$ might under-estimate the benefit conferred by these interventions to treated individuals. 

%Therefore the risk-benefit trade-offs for those who are untreated are unknown. }

When might the individual effect $RD^I$ be of scientific interest? First, resource constraints might necessitate a policy in which a single subject (or a fixed number of subjects) is treated per cluster, so a trial design that enforces this constraint may naturally reveal the quantity of interest.  Second, $RD^I$ may be of interest to investigators and research subjects because it summarizes the ethical trade-off in benefit or harm experienced by treated versus untreated individuals. 
In the case of PrEP, these trade-offs might involve the benefit of reduced HIV infection risk for all cluster members, versus potential medication side effects for individuals treated with PrEP \mbox{\citep{thigpen2012antiretroviral}}. 
That is, $RD^I$ answers the question, ``am I better off being treated or untreated, when someone in my cluster is treated?'' 
If investigators desire a measure of the effect of an intervention on an individual subject while holding disseminated exposure constant, the ``composite'' effect $RD^{\text{Comp}}$ may be a more readily interpretable causal estimand.

%%%%%%% ACKNOWLEDGEMENTS %%%%%%%%%%%%%%%%%%%%%%%%

\textbf{Acknowledgements:} We are grateful to 
Peter M. Aronow, 
Samuel R. Friedman,
Gregg S. Gonsalves,
M. Elizabeth Halloran, 
Joseph Lewnard, and 
Fredrik S\"avje 
for helpful comments.  We acknowledge funding from NIH grants NICHD DP2 HD0917991, NIDA R36 DA042643, NIEHS DP1 ES025459, NIDA 1DP2 DA046856-01, NIAID 5R01 AI112339-02, and NIGMS U54 GM115677.

%%%%%%%% REFERENCES %%%%%%%%%%%%%%%%%%%%%%%%%%%%%

\setlength{\bibsep}{1.0pt}
\bibliographystyle{plainnat}
\bibliography{network_causal_effects}

\begin{thebibliography}{15}
\providecommand{\natexlab}[1]{#1}
\providecommand{\url}[1]{\texttt{#1}}
\expandafter\ifx\csname urlstyle\endcsname\relax
  \providecommand{\doi}[1]{doi: #1}\else
  \providecommand{\doi}{doi: \begingroup \urlstyle{rm}\Url}\fi

\bibitem[Baeten et~al.(2012)Baeten, Donnell, Ndase, Mugo, Campbell, Wangisi,
  Tappero, Bukusi, Cohen, Katabira, et~al.]{baeten2012antiretroviral}
Jared~M Baeten, Deborah Donnell, Patrick Ndase, Nelly~R Mugo, James~D Campbell,
  Jonathan Wangisi, Jordan~W Tappero, Elizabeth~A Bukusi, Craig~R Cohen, Elly
  Katabira, et~al.
\newblock Antiretroviral prophylaxis for {HIV} prevention in heterosexual men
  and women.
\newblock \emph{New England Journal of Medicine}, 367\penalty0 (5):\penalty0
  399--410, 2012.

\bibitem[Benjamin-Chung et~al.(2017)Benjamin-Chung, Arnold, Berger, Luby,
  Miguel, Colford~Jr, and Hubbard]{benjamin2017spillover}
Jade Benjamin-Chung, Benjamin~F Arnold, David Berger, Stephen~P Luby, Edward
  Miguel, John~M Colford~Jr, and Alan~E Hubbard.
\newblock Spillover effects in epidemiology: parameters, study designs and
  methodological considerations.
\newblock \emph{International Journal of Epidemiology}, 47\penalty0
  (1):\penalty0 332--347, 2017.

\bibitem[Buchanan et~al.(2018)Buchanan, Vermund, Friedman, and
  Spiegelman]{buchanan2018assessing}
Ashley~L Buchanan, Sten~H Vermund, Samuel~R Friedman, and Donna Spiegelman.
\newblock Assessing individual and disseminated effects in network-randomized
  studies.
\newblock \emph{American Journal of Epidemiology}, 187\penalty0 (11):\penalty0
  2449--2459, 2018.

\bibitem[Choopanya et~al.(2013)Choopanya, Martin, Suntharasamai, Sangkum, Mock,
  Leethochawalit, Chiamwongpaet, Kitisin, Natrujirote, Kittimunkong,
  et~al.]{choopanya2013antiretroviral}
Kachit Choopanya, Michael Martin, Pravan Suntharasamai, Udomsak Sangkum,
  Philip~A Mock, Manoj Leethochawalit, Sithisat Chiamwongpaet, Praphan Kitisin,
  Pitinan Natrujirote, Somyot Kittimunkong, et~al.
\newblock Antiretroviral prophylaxis for {HIV} infection in injecting drug
  users in {B}angkok, {T}hailand (the {B}angkok {T}enofovir {S}tudy): a
  randomised, double-blind, placebo-controlled phase 3 trial.
\newblock \emph{The Lancet}, 381\penalty0 (9883):\penalty0 2083--2090, 2013.

\bibitem[Cohen et~al.(2011)Cohen, Chen, McCauley, Gamble, Hosseinipour,
  Kumarasamy, Hakim, Kumwenda, Grinsztejn, Pilotto,
  et~al.]{cohen2011prevention}
Myron~S Cohen, Ying~Q Chen, Marybeth McCauley, Theresa Gamble, Mina~C
  Hosseinipour, Nagalingeswaran Kumarasamy, James~G Hakim, Johnstone Kumwenda,
  Beatriz Grinsztejn, Jose~HS Pilotto, et~al.
\newblock Prevention of {HIV}-1 infection with early antiretroviral therapy.
\newblock \emph{New England Journal of Medicine}, 365\penalty0 (6):\penalty0
  493--505, 2011.

\bibitem[Donnell et~al.(2010)Donnell, Baeten, Kiarie, Thomas, Stevens, Cohen,
  McIntyre, Lingappa, Celum, in~Prevention HSV/HIV Transmission Study~Team,
  et~al.]{donnell2010heterosexual}
Deborah Donnell, Jared~M Baeten, James Kiarie, Katherine~K Thomas, Wendy
  Stevens, Craig~R Cohen, James McIntyre, Jairam~R Lingappa, Connie Celum,
  Partners in~Prevention HSV/HIV Transmission Study~Team, et~al.
\newblock Heterosexual {HIV}-1 transmission after initiation of antiretroviral
  therapy: a prospective cohort analysis.
\newblock \emph{The Lancet}, 375\penalty0 (9731):\penalty0 2092--2098, 2010.

\bibitem[Eck et~al.(2018)Eck, Morozova, and Crawford]{eck2018randomization}
Daniel~J. Eck, Olga Morozova, and Forrest~W. Crawford.
\newblock Randomization for the direct effect of an infectious disease
  intervention in a clustered study population.
\newblock \emph{arXiv:1808.05593}, 2018.

\bibitem[Green et~al.(2014)Green, Ray, Bowman, McKenzie, and
  Rich]{green2014two}
Traci~C Green, Madeline Ray, Sarah~E Bowman, Michelle McKenzie, and Josiah~D
  Rich.
\newblock Two cases of intranasal naloxone self-administration in opioid
  overdose.
\newblock \emph{Substance Abuse}, 35\penalty0 (2):\penalty0 129--132, 2014.

\bibitem[Group(2015{\natexlab{a}})]{insight2015initiation}
Insight Start~Study Group.
\newblock Initiation of antiretroviral therapy in early asymptomatic {HIV}
  infection.
\newblock \emph{New England Journal of Medicine}, 373\penalty0 (9):\penalty0
  795--807, 2015{\natexlab{a}}.

\bibitem[Group(2015{\natexlab{b}})]{temprano2015trial}
Temprano ANRS 12136~Study Group.
\newblock A trial of early antiretrovirals and isoniazid preventive therapy in
  {A}frica.
\newblock \emph{New England Journal of Medicine}, 373\penalty0 (9):\penalty0
  808--822, 2015{\natexlab{b}}.

\bibitem[Hudgens and Halloran(2008)]{hudgens2008toward}
Michael~G Hudgens and M~Elizabeth Halloran.
\newblock Toward causal inference with interference.
\newblock \emph{Journal of the American Statistical Association}, 103\penalty0
  (482):\penalty0 832--842, 2008.

\bibitem[Organization(2016)]{world2016consolidated}
World~Health Organization.
\newblock \emph{Consolidated guidelines on the use of antiretroviral drugs for
  treating and preventing {HIV} infection: recommendations for a public health
  approach}.
\newblock World Health Organization, 2016.

\bibitem[S{\"a}vje et~al.(2017)S{\"a}vje, Aronow, and
  Hudgens]{savje2017average}
Fredrik S{\"a}vje, Peter~M Aronow, and Michael~G Hudgens.
\newblock Average treatment effects in the presence of unknown interference.
\newblock \emph{arXiv:1711.06399}, 2017.

\bibitem[Thigpen et~al.(2012)Thigpen, Kebaabetswe, Paxton, Smith, Rose,
  Segolodi, Henderson, Pathak, Soud, Chillag,
  et~al.]{thigpen2012antiretroviral}
Michael~C Thigpen, Poloko~M Kebaabetswe, Lynn~A Paxton, Dawn~K Smith, Charles~E
  Rose, Tebogo~M Segolodi, Faith~L Henderson, Sonal~R Pathak, Fatma~A Soud,
  Kata~L Chillag, et~al.
\newblock Antiretroviral preexposure prophylaxis for heterosexual {HIV}
  transmission in {B}otswana.
\newblock \emph{New England Journal of Medicine}, 367\penalty0 (5):\penalty0
  423--434, 2012.

\bibitem[VanderWeele and Tchetgen(2011)]{vanderweele2011effect}
Tyler~J VanderWeele and Eric J~Tchetgen Tchetgen.
\newblock Effect partitioning under interference in two-stage randomized
  vaccine trials.
\newblock \emph{Statistics \& Probability Letters}, 81\penalty0 (7):\penalty0
  861--869, 2011.

\end{thebibliography}

\end{document}